\documentclass[manuscript]{rmaa}

\usepackage{paralist}

\usepackage{psfrag,color}

\usepackage[latin1]{inputenc}

\usepackage{lscape}
\usepackage{graphicx}
\usepackage{color}
\usepackage{xcolor}
\usepackage{ulem}
\usepackage{txfonts}
\usepackage{natbib}
\bibpunct{(}{)}{;}{a}{}{,}

\newcommand{\cmmth}{\mbox{cm$^{-3}$}}

\newcommand{\kms}{\mbox{km\,s$^{-1}$}}

\newcommand{\mjyb}{\mbox{mJy\,beam$^{-1}$}}
\newcommand{\mujyb}{\mbox{$\mu$Jy\,beam$^{-1}$}}
\newcommand{\masyr}{\mbox{mas\,yr$^{-1}$}}
\newcommand{\msun}{\mbox{$M_{\odot}$}}

\newcommand\degr{\mbox{$^\circ$}}%

\title{A one-sided knot ejection at the core of the  HH~111 outflow} 

\author{
L. G\'omez,\altaffilmark{1,2,3}
L. F. Rodr\'\i guez,\altaffilmark{4,5}
L. Loinard\altaffilmark{1,4}
}

\altaffiltext{1}{Max-Planck-Institut f\"ur Radioastronomie, Germany.}
\altaffiltext{2}{Departamento de Astronom\'\i{}a, Universidad de Chile,
Chile.}
\altaffiltext{3}{CSIRO Astronomy and Space Science, Australia.}
\altaffiltext{4}{Centro de Radioastronom\'\i{}a y Astrof\'\i{}sica,
  UNAM, M\'exico.}
\altaffiltext{5}{Astronomy Department, Faculty of Science, King Abdulaziz 
             University, Saudi Arabia.}

\shortauthor{G\'omez, Rodr\'\i guez, \& Loinard}
\shorttitle{A one-sided knot ejection in HH 111}

\fulladdresses{
\item Laura G\'omez: Departamento de Astronom\'\i{}a, Universidad de Chile,
      Camino El Observatorio 1515, Las Condes, Santiago, Chile (lgomez@das.uchile.cl).
\item Laurent Loinard and Luis F. Rodr\'\i guez: Centro de Radioastronom\'\i{}a
      y Astrof\'\i{}sica, Universidad Nacional Aut\'onoma de M\'exico,
      Apartado Postal 3-72, 58090 Morelia, Michoac\'an, M\'exico
      (l.loinard, l.rodriguez@crya.unam.mx).
}

\listofauthors{L. G\'omez,  L. F. Rodr\'\i guez, \& L. Loinard}
\indexauthor{G\'omez, L.}
\indexauthor{Rodr\'\i guez, L. F.}
\indexauthor{Loinard, L.}

\abstract{
We present an astrometry study of the radio source VLA~1 at the core of the 
HH~111 outflow using new data (2007) as well as archival observations
(1992-1996). All data were taken at 3.6~cm with the Very Large Array 
in its most extended (A) configuration.
The source VLA~1 has undergone a dramatic morphological change, showing a one-sided
knot ejection in the 2007 epoch. We also report on the detection of a 3.6~cm compact 
continuum source (VLA~3) located at ($-$10\farcs6,\,98\farcs7) from VLA~1. No significant absolute proper motions were found for VLA~1 and VLA~3
and the upper limits are consistent with those found for (embedded) radio sources in the 
Orion Nebula.
We favor the interpretation that in the
continuum at 3.6~cm we
are observing two nearly perpendicular jets. 
HH~111 presents a new case of one-sided jet ejection in a young stellar object. 
The Galactic (or extragalactic) nature of VLA~3 remains unclear.
}

\resumen{
Presentamos un estudio astrom\'etrico
de la radiofuente VLA~1 de donde se origina el flujo HH~111 usando nuevos datos
(2007) as\'\i ~como tambi\'en observaciones de archivo (1992-1996). Todos los
datos fueron tomados a 3.6 cm con el Very Large Array en su configuraci\'on 
m\'as extendida (A).
La fuente VLA~1 ha tenido un cambio dram\'atico, mostrando una eyecci\'on  
monopolar en la \'epoca de 2007. Tambi\'en reportamos la detecci\'on de una 
fuente
de continuo compacta (VLA~3) localizada a ($-$10\farcs6,\,98\farcs7) de VLA~1.
No se encontraron movimientos propios significativos en VLA~1 ni en VLA~3 y los
l\'\i mites superiores son consistentes con los que se encontraron para
radiofuentes (embebidas) en la Nebulosa de Ori\'on.
Favorecemos la interpretaci\'on de que a 3.6~cm estamos observando dos chorros
casi perpendiculares. HH~111 representa un nuevo caso de eyecciones monopolares 
en un objeto joven estelar. No se pudo establecer si VLA~3 es Gal\'actica (o extragal\'actica).
}

\addkeyword{Astrometry}
\addkeyword{ISM: individual objects: HH~111}
\addkeyword{ISM: jets and outflows}
\addkeyword{radio continuum: ISM}

\begin{document}

\maketitle

\section{Introduction}

Many observations have increasingly shown that stars are rarely born alone
\citep[e.g.,][]{la03}. 
Binary and multiple systems may be surrounded by circumstellar and/or
circumbinary disks, and often drive episodic jets and powerful outflows.  
For instance, \citet{pe10} presented Very Large Array (VLA) multi-epoch 
observations of the 
very young hierarchical multiple system IRAS~16293$-$2422. These authors
found a bipolar ejection and suggested it was a consequence of an episode of 
enhanced mass loss
possibly produced by an increase in accretion onto the protostar.

An interesting binary system is found at the core of the giant Herbig-Haro 
object HH~111, originally discovered by \citet{re89}. 
This flow in Orion has been extensively studied
since its discovery \citep[see][and references therein]{no11}. 
It is highly collimated approximately in the east-west
direction and displays a large number of individual knots 
moving in the plane of the sky with velocities of several hundred 
\kms~\citep{re92}. We will refer to this east-west jet as the
HH~111 optical jet.
Near-infrared observations by \citet{gr93}
revealed the presence of a second outflow (HH~121) nearly perpendicular to 
the HH~111 one, making the system a quadrupolar outflow.
We will refer to this second jet, approximately in the north-south
direction, as the infrared jet.

The driving sources of the HH~111 jets are believed to be the components of the
suspected binary, class I source IRAS~05491+0247, which was associated
to the radio source VLA~1 reported by \citet{re99} at 3.6~cm. These authors also
detected an additional faint, unresolved, radio source (VLA~2) to the 
north-west of VLA~1. 
Recently, \citet{lee11} imaged the HH~111 protostellar system at 1.3~mm.
A faint (4$\sigma$) source appeared at the same position 
of VLA~2 possibly tracing its putative dusty disk.

\citet{ro08} presented observations of the system at 7~mm and discussed three 
possible interpretations for VLA~1, since the emission revealed a 
structure that can be described as
two overlapping, almost perpendicular, elongated sources.
They favor two interpretations in which the
quadrupolar structure was either a disk elongated in the
north-south direction with a perpendicular jet or two orthogonal disks.
In the third interpretation, the emission comes from two jets. Nevertheless,
the free-free contribution is only about 30\%, hence the emission is dominated by 
dust. However, it is possible that the core is dominated by dust and the extended
emission dominated by free-free emission \citep[see][]{ro08}.
 The emission at 1.3~mm that showed a structure elongated in the north-south 
direction and perpendicular to the HH~111 optical jet  \citep{re99}
 was attributed to the thermal emission from a disk \citep{lee11}.

In this paper we present sensitive, high angular-resolution 
3.6~cm continuum observations at the core of the HH~111 outflow
to search for absolute proper motions and to study the evolution of the
HH~111 jet in VLA~1. In Sect.~\ref{obs-datared}, we 
describe our observations carried out with the Very Large Array as well as the 
data reduction. In Sect.~\ref{results}, we present the results together with 
the analysis. The discussion is presented in Sect.~\ref{hh111:discussion} 
in the context of one-sided knot ejections versus bipolar ejections
and the summary is given in Sect.~\ref{hh111:summary}.

\begin{table*}
\centering
\small
\begin{changemargin}{-2cm}{-2cm}
\caption{Very Large Array observations at 3.6~cm.}
\label{Tepochs}
\setlength{\tabnotewidth}{0.90\linewidth}
\tablecols{6}
 \begin{tabular}{cccccc}
\toprule \toprule

Epoch & Project & Phase calibrator & Bootstrapped flux density  & Synthesized beam\tabnotemark{a}  & $\sigma$ \\

      &         &      &   (Jy) &  & (\mjyb)\\
\midrule

1992 Nov \phantom{0}2 (1992.84) & AR278 & 0541$-$056 & $ 0.8691 \pm 0.0002$ & $0\farcs26 \times 0\farcs23; +42\degr$ &  0.02  \\
1992 Dec 18 (1992.96) & AR278 & 0541$-$056 & $1.0253 \pm 0.0002$ & $0\farcs25 \times 0\farcs24; +37\degr$ &   0.02  \\
1992 Dec 19 (1992.97) & AR278 & 0541$-$056 & $1.0229 \pm 0.0002$ & $0\farcs28 \times 0\farcs25; +8\degr$ &   0.02   \\
1994 Apr 30 (1994.33) & AR277 &  0552+032  & $0.7663 \pm 0.0003$ & $0\farcs21 \times 0\farcs20; -57\degr$ &  0.02   \\
1996 Nov 11 (1996.86) & AR367 & 0552+032\tabnotemark{b} & $0.8085 \pm  0.0001$ & $0\farcs25 \times 0\farcs22; -12\degr$ &   0.01  \\
1996 Dec 28 (1996.99) & AR277 & 0552+032 & $0.6997 \pm  0.0001$ & $0\farcs25 \times 0\farcs22; -5\degr$ &  0.02  \\
1996 Dec 29 (1997.00) & AR277 & 0552+032 & $0.7088 \pm  0.0001$ & $0\farcs30 \times 0\farcs25; +3\degr$ &   0.02     \\
2007 Aug 12 (2007.61) & AG747 &  0552+032  & $0.6780 \pm  0.0003$ & $0\farcs26 \times 0\farcs23; -11\degr$ &  0.01  \\

\bottomrule

\tabnotetext{a}{\small Major axis $\times$ minor axis; position angle of major axis.}
\tabnotetext{b}{\small For this epoch, the source 0541$-$056 was also observed.}

\end{tabular}
\end{changemargin}
\end{table*}


\section{Observations and data reduction}\label{obs-datared}

The 3.6~cm ($\nu$ = 8.4~GHz) continuum observations of the core of the HH~111 
outflow were made with the Very Large Array (VLA) of the National Radio 
Astronomy Observatory (NRAO\footnote{The National Radio Astronomy Observatory 
is a facility of the National Science Foundation operated under cooperative 
agreement by Associated Universities, Inc.}) in the A configuration on 
2007 August 12. The phase center of the observations was at  
$\alpha$(J2000)~=~5$^{\rm h}$51$^{\rm m}$46\fs25, 
$\delta$(J2000)~=~2\degr48\arcmin29\farcs6. 
The phase calibrator 0541$-$056 was observed for the first three 1992 epochs
while the source 0552+032 was used as phase calibrator for the rest of the 
epochs (see Table~\ref{Tepochs}). 
The absolute amplitude calibrator was the source 1331+305. We used
a model image for 1331+305 provided by NRAO in order to improve flux accuracy.

In addition, we used seven epochs taken from the VLA archive. The eight
analyzed data sets span about 15 years and are listed in Table~\ref{Tepochs}.
The data were analyzed in the standard manner using the 
Astronomical Image Processing System (AIPS) software package;
the calibrated visibilities were imaged using weights intermediate
between natural and uniform (with the ROBUST parameter set to 0) and 
{\sc CLEAN}ed in an iterative fashion.
To obtain accurate absolute astrometry, we precessed the data taken in B1950.0 
to J2000.0 using the task {\sc uvfix} and used the most recent
position of the phase calibrator for all epochs. We expect that the residual
systematic error affecting the data to be around 10 milliarcseconds (mas).
These residual errors were added in quadrature to the positional uncertainty
delivered by a Gaussian fitting program (task {\sc jmfit} of AIPS).


\section{Results and analysis}\label{results}

\begin{figure}
	  \centering
	  \includegraphics[angle=-90,width=8.5cm]{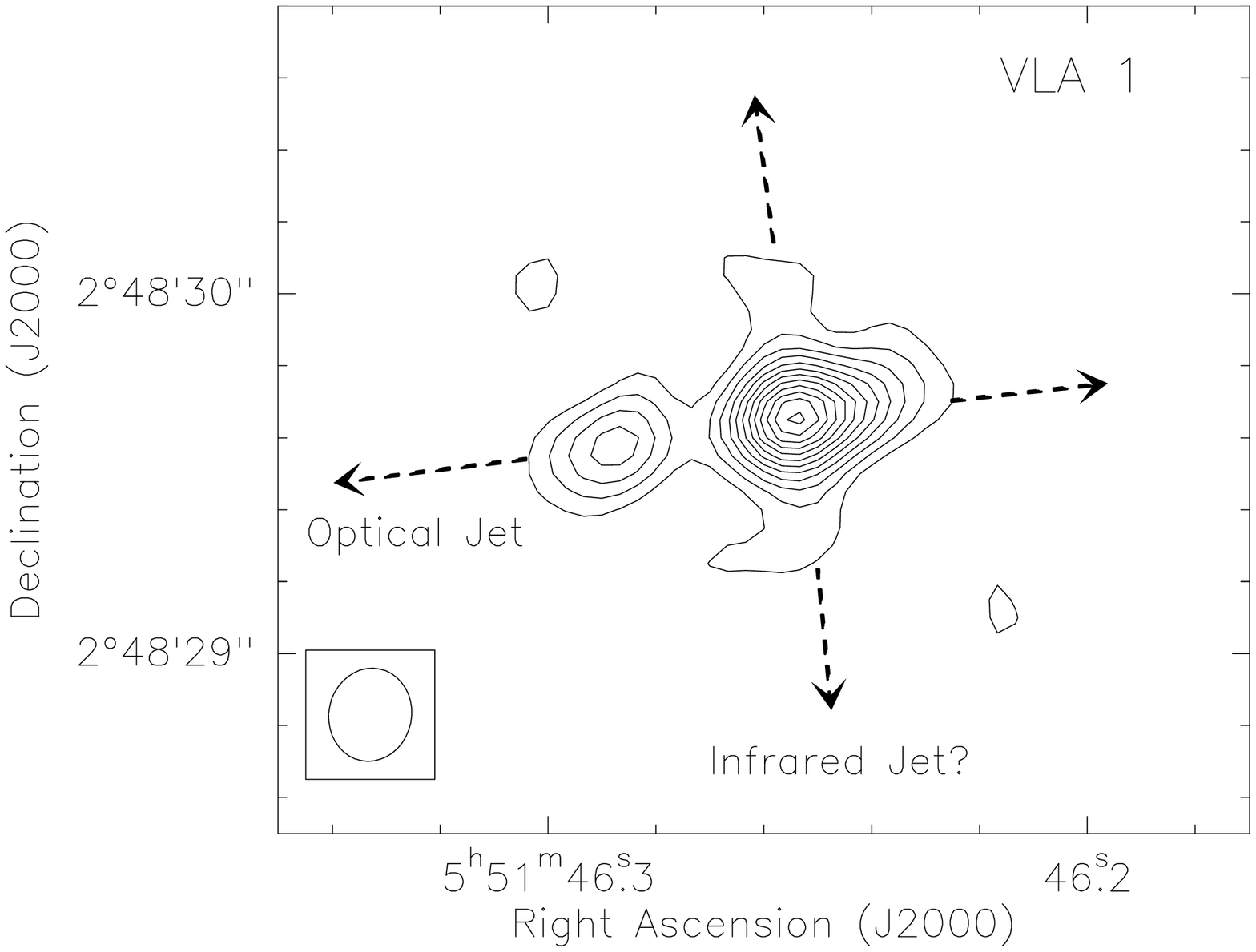}\\
	  \vspace{0.3cm}
	      \includegraphics[angle=-90,width=8.5cm]{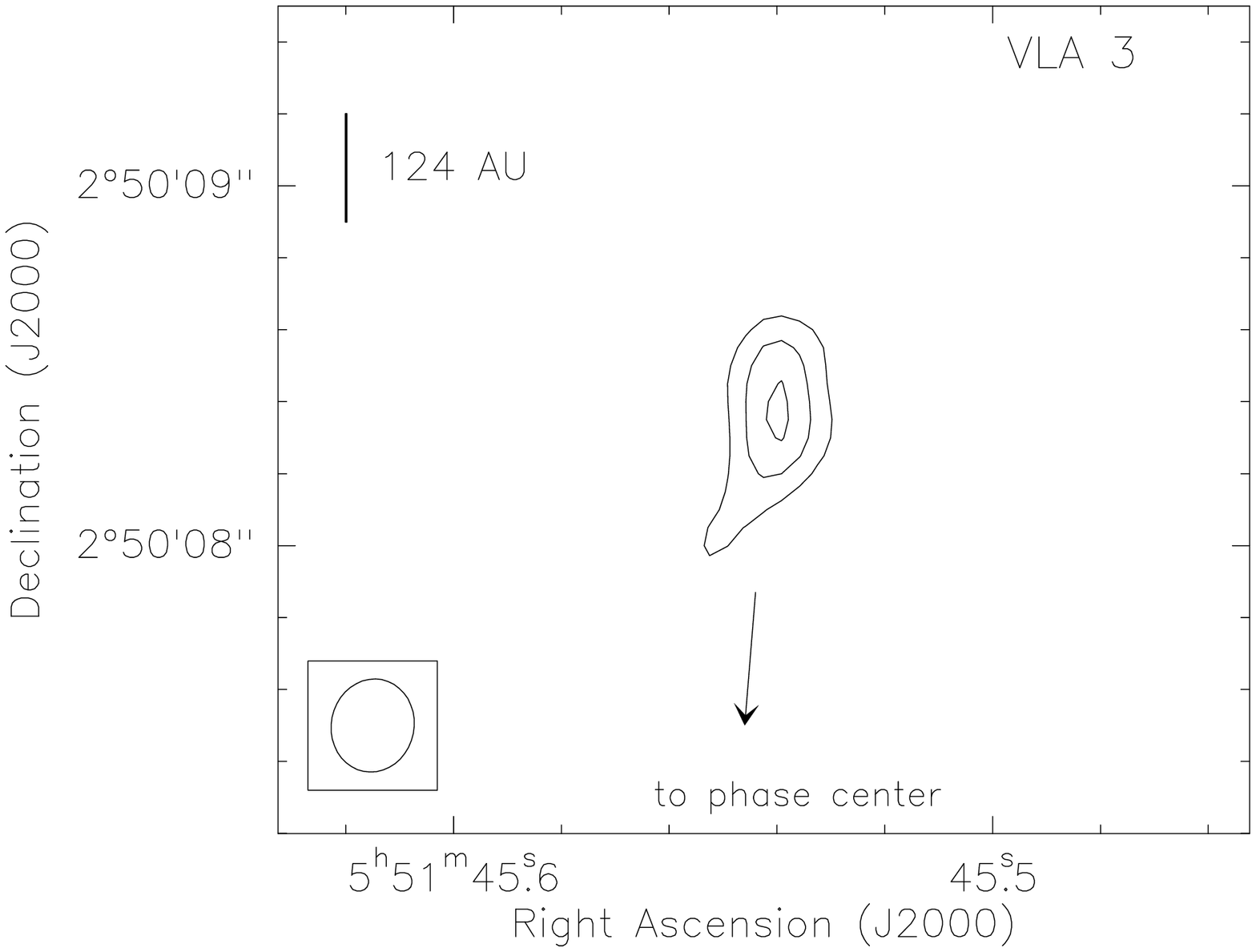}
		     \caption{
Contour images of the 3.6~cm continuum emission of VLA~1 ({\it top})
and VLA~3 ({\it bottom}) at epoch 2007.61. The
synthesized beam ($0\farcs26 \times 0\farcs23$; PA = $-11\degr$) is 
shown
in the bottom left corner.
{\it Top}: First contour and contour spacing for VLA~1 are
31.5~\mujyb~(3$\sigma$). The arrows in the east-west direction indicate 
the
P.A. of the HH~111 jet~\citep[97.5\degr;][]{no11}. The arrows in the
north-south direction indicate the orientation of the HH~121 infrared jet.
 {\it Bottom}: First contour and contour spacing for VLA~3 are
 42.0~\mujyb~(3$\sigma$). The bar indicates the linear scale. The arrow 
 points in the direction to the
 phase center. The map has been corrected for the primary beam response.
				     }
					       \label{Fhh111-n-v1}
						   \end{figure}

We detected the source VLA~1 at epoch 2007.61 and show the image in the 
{\it upper} panel of Fig.~\ref{Fhh111-n-v1}.
The morphology of VLA~1 changed drastically between the early epochs \citep{ro94,re99} 
and the 2007.61 epoch. In the last epoch
there is a knot ejection to the east which was not
present in previous observations. Parameters of
the ejecta and discussion of the one-sided ejection are presented in
Sects.~\ref{parameters:v1} and~\ref{discussion:v1}, respectively.
Weak emission (at 3$\sigma$ level) to the north and south of VLA~1 
is also detected. We believe this weak emission is associated with the
base of the infrared jet.

In the 2007 observations, due to potential 
problems with several antennas that were in transition to the K. Jansky Very 
Large Array (VLA) system, the rms noise level was not sufficient 
to detect the VLA~2 source, as previously reported in \citet{re99}.

      \begin{figure}
   \centering
   \includegraphics[angle=-90,width=15.5cm]{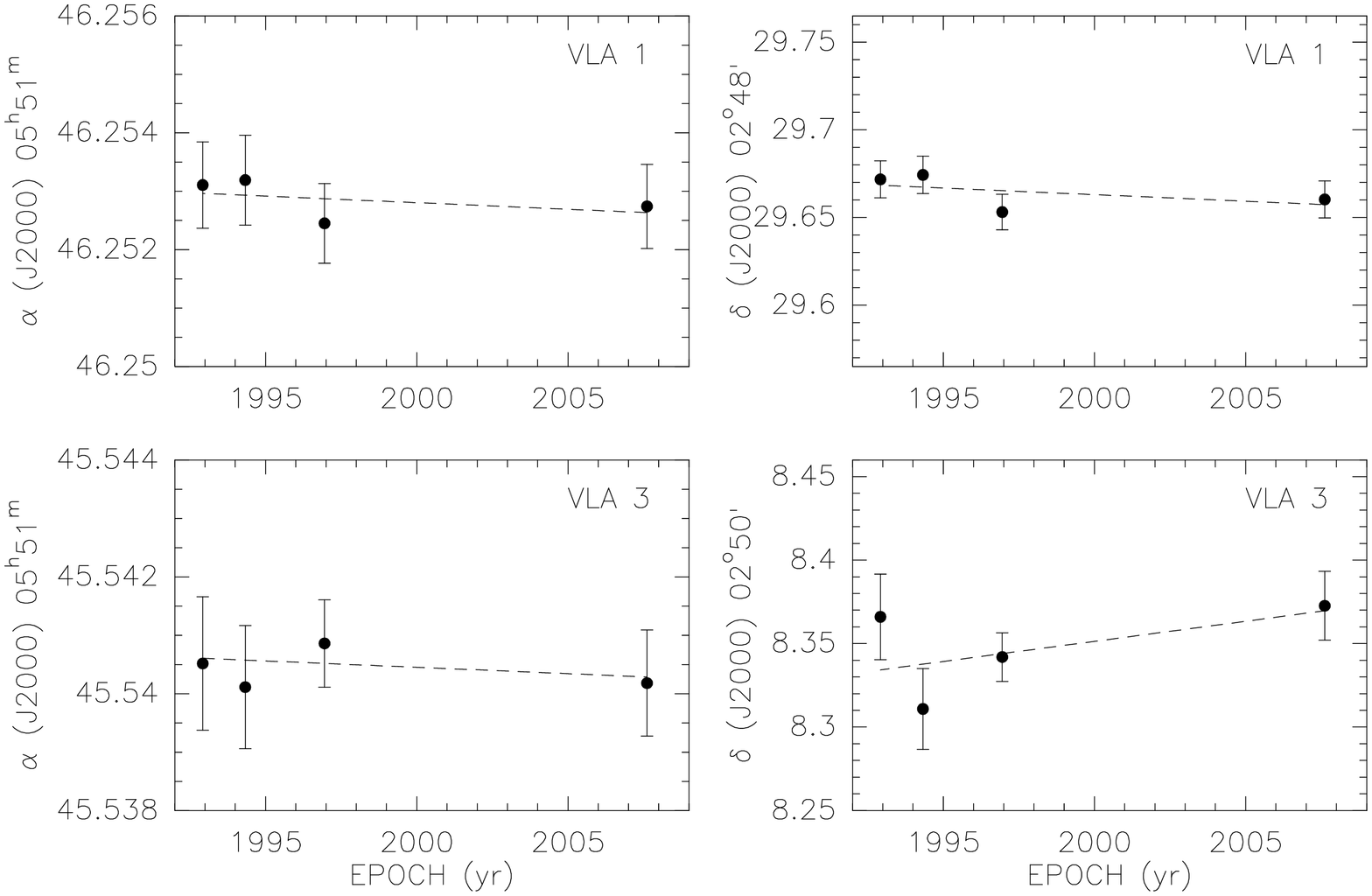}
      \caption{Absolute proper motions in right ascension ($\alpha$) and in
declination ($\delta$) for VLA~1 ({\it upper panels}) and the newly reported 
radio
source VLA~3 ({\it bottom panels}) as a function of time. The dashed line
in each panel represents the least squares fit to the data.
The right ascension axis is given in seconds, while the declination axis is
given in arcseconds.
               }
         \label{Fproperm}
   \end{figure}

On the other hand, we report on the detection of a radio source 
(hereafter VLA~3) located at 
($-$10\farcs6,\,98\farcs7) from VLA~1.  
Inspection of the maps at other epochs confirms the presence of this source. 
After correcting the image for the primary beam response, 
VLA~3 has a flux density of $S = 0.22 \pm 0.03$~mJy in the 2007 image.
The flux density at all four epochs is consistent with a value
of $S = 0.3 \pm 0.1$~mJy. There is a suggestion of time variability,
but the data of 1992 and 1994 do not have a sufficient 
signal-to-noise ratio to reach a firm conclusion.
The elongation of the source in the north-south direction 
seen in the {\it bottom} panel of Fig.~\ref{Fhh111-n-v1} is caused by 
bandwidth smearing and is not intrinsic to the source.

\subsection{Proper motions}\label{prop-mot}

In order to search for absolute proper motions of VLA~1 and VLA~3, 
we concatenated the epochs 1992.84, 1992.96, and 1992.97 into epoch 1992.92;
and the epochs 1996.86, 1996.99, and 1997.00 into epoch 1996.95. 
Figure~\ref{Fproperm} shows the radio positions of VLA~1 and VLA~3 
as a function of time. We performed a least squares fit to four epochs
(1992.92, 1994.33, 1996.95, and 2007.61) and the results are listed in Table~\ref{Tpropm}.
We found that VLA~1 and VLA~3 do not show significant absolute  proper motions 
and they are consistent within 2$\sigma$ with those found for (embedded) radio sources in the Orion Nebula 
\citep{lau05}.

\begin{table}
\caption{Absolute proper motions.}
\label{Tpropm}
\centering
\begin{tabular}{l c c} 
\toprule \toprule
Source &  $\mu_\alpha \cos\delta$  & $\mu_\delta$   \\
       &        (\masyr)          & (\masyr)    \\
\midrule
VLA 1  & $-0.34 \pm 0.50$  &  $-0.75 \pm 0.93$  \\
VLA 3  & $-0.33 \pm 0.57$  &  $\phantom{0}2.41 \pm 2.23$  \\
\bottomrule
\end{tabular}
\end{table}

Further observations will be necessary to detect VLA~2,
so that we are able to obtain relative proper motions with respect to VLA~1 and 
test the hypothesis that these sources form a non-hierarchical triple system 
in disintegration \citep{re99}.

\subsection{On the nature of VLA~3}\label{nature:vla3}

In order to obtain the expected number of extragalactic sources, $N$, in the 
field of study with a flux density comparable or larger
than that of VLA~3, we follow \citet{wi93}
\begin{equation}
N = 2.4\times 10^{-3}~\left(\frac{S}{\rm{mJy}}\right)^{-1.3}\label{eextra}\,~{\rm arcmin}^{-2}\,,
\end{equation}
where $S$ is the flux density in mJy. For $S \simeq 0.3$~mJy and a
field of  $\sim 3.4 \times 3.4$~arcmin$^{2}$, $N$ corresponds to 0.13 sources.
Thus, VLA~3 only has a 13\% probability of being extragalactic.

VLA~3 does not show significant proper motions, as expected for an
extragalactic source, but is also consistent with Orion sources.
Therefore, the measurement of proper motions, in this case, is not 
useful in determining its nature.
Very recently, the
\citet{ai12} detected a radio source whose emission peaks  
at the position of VLA~3 at 1.9~cm (16~GHz) with a beam size of 
$69\arcsec \times 27\arcsec$.
They obtained a spectral index, $\alpha_{\rm AMI}$, of  $-0.23 \pm 1.10$, a value 
consistent with those found for extragalactic sources, but whose large
error makes it consistent with several different emission mechanisms. 

We searched for possible counterparts in three point source catalogs, 
namely, the AKARI \citep{mu07}, 2MASS \citep{sk06}, and WISE \citep{wi10} catalogs.
We found no association with sources from the AKARI and 2MASS catalogs.
A WISE point source was found as possible counterpart of VLA~3 at
a distance of $\sim$1\arcsec. The source is detected in the 3.4~$\mu$m and 
4.6~$\mu$m bands. There are marginal signals in  the 12~$\mu$m and 22~$\mu$m 
bands that are considered as spurious or affected by other 
artifacts in the catalog.

Although VLA~3 is more likely an extragalactic source, in the 
remaining of this contribution we restrain ourselves from further 
interpretation of VLA~3.

\subsection{VLA~1: ejecta parameters}\label{parameters:v1}

The parameters of the ejecta are estimated from observables at epoch 2007.61.
Using the model by \citet{me67}, and assuming optically 
thin free-free emission, the mass of ionized gas, $M_i$, and the electron 
density, $n_e$, of a homogeneous sphere can be calculated as
\begin{eqnarray}
M_i & = & 3.39~\times~10^{-5}~\left(\frac{T_e}{10^4~{\rm K}}\right)^{0.175}\left(\frac{\nu}{1~{\rm GHz}}\right)^{0.05}\left(\frac{S_{\nu}}{1~{\rm mJy}}\right)^{0.5}
\nonumber \\
& & {}\times \left(\frac{D}{1~{\rm kpc}}\right)^{2.5}\left(\frac{\theta}{1~\arcsec}\right)^{1.5} \label{eimass}\,~\msun,\\
n_e & = & 7.2~\times~10^{3}~~\left(\frac{T_e}{10^4~{\rm K}}\right)^{0.175}\left(\frac{\nu}{1~{\rm GHz}}\right)^{0.05}\left(\frac{S_{\nu}}{1~{\rm mJy}}\right)^{0.5}
\nonumber \\
& & {}\times \left(\frac{D}{1~{\rm kpc}}\right)^{-0.5}\left(\frac{\theta}{1~\arcsec}\right)^{-1.5}\label{ene}\,~\cmmth,
\end{eqnarray}
where $T_e$ is the electron temperature, $\nu$ is the frequency, $S_{\nu}$ is 
the observed integrated flux density, $D$ is the distance, and
$\theta$ is the angular size. We assume a $T_e$ of 10$^4$~K at 8.4~GHz,
and use a distance of 414 pc \citep{me07}. The
flux density was obtained by fitting a Gaussian to the ejecta. 
The emission was resolved in only one direction with 
a deconvolved angular size of 0\farcs28,  
while in the other direction the emission comes from a region smaller than
0\farcs23.
Using the fitting parameters of $S = $~$0.18$~mJy and the geometric mean of
the deconvolved angular size of
$\lesssim 0\farcs25$, we obtain $M_i \lesssim 2.2 \times 10^{-7}$~\msun~and
$n_e \gtrsim 4.2 \times 10^{4}$ \cmmth.


\section{Discussion}\label{hh111:discussion}

\subsection{A one-sided knot ejection}\label{discussion:v1}

\begin{figure}
   \centering
   \includegraphics[trim=0cm 0.7cm 0cm 0.7cm,clip=true,width=6cm]{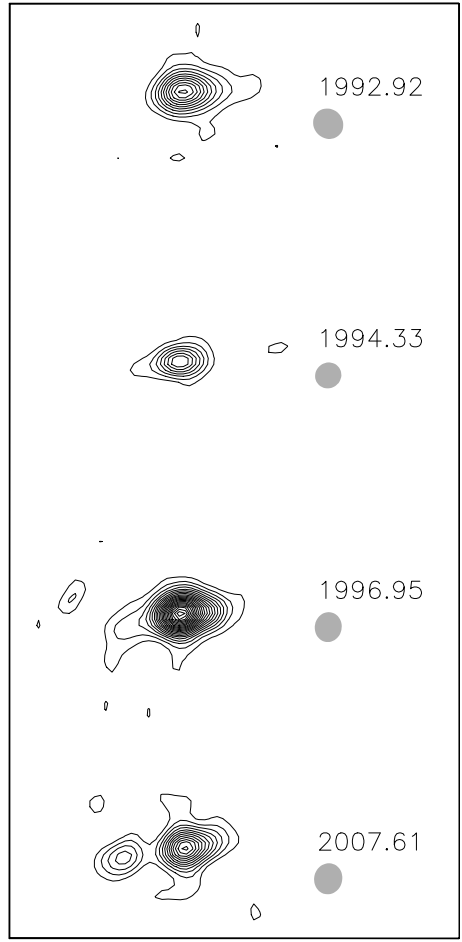}
      \caption{
Contour images of the 3.6~cm continuum emission of VLA~1 at
different epochs. Details for epoch 2007.61 are the same as in 
Fig.~\ref{Fhh111-n-v1}.
First contour and contour spacing for VLA~1 are, from {\it top} to {\it bottom},
36.0~\mujyb~(3$\sigma$), 54.3~\mujyb~(3$\sigma$), and 21.0~\mujyb~(3$\sigma$), 
for epoch 1992.92, 1994.33, and 1996.95, respectively. The synthesized beams  are shown
as filled ellipses for each epoch [($0\farcs26 \times 0\farcs24$; PA~=~$34\degr$), ($0\farcs22 \times 0\farcs21$; PA = $-41\degr$), ($0\farcs25 \times 0\farcs22$; PA = $-8\degr$)].
               }
         \label{evolution}
   \end{figure}

Figure~\ref{evolution} compares the images for VLA~1 of four epochs, 
including the 2007 observations. We show images of concatenated data for
epochs 1992 and 1996 (see Sect.~\ref{prop-mot}).
For the 1992 epoch, \citet{ro94} note that the emission of VLA~1 consists of
a single component elongated to the west, possibly tracing a faint ejection. The elongation is to the 
east in the 1994 epoch, and to the east-west direction for the structure
seen in the 1996 epoch,
 whereas in the new 
2007 observations, we additionally see a second component to the 
east, which we interpret as a one-sided knot ejection 
(see also Fig.~\ref{Fhh111-n-v1}). Moreover, weak emission is detected in the
north-south direction, and according to \citet{ro08}, it probably is due to the presence of the 
second (infrared) HH~121 jet. In the 3.6~cm map presented by \citet{re99}, VLA~1 
shows clear elongations in the north-south direction after using a maximum
entropy reconstruction (see their Fig.~3).

It is important to point out that the map corresponding to the 1996
epoch, which has a better rms compared to the 2007 epoch, presents an
elongation to the east but that the ejection cannot be resolved.
It is also interesting that, in the optical, the (eastern) redshifted lobe is 
completely obscured \citep[see][]{re92}, contrary to what we see at radio 
wavelengths in the 2007 epoch where the ejection is clearly seen to the east.

Dust extinction affects observations at 
optical and infrared wavelengths, while the radio regime suffers the least. 
One may think that asymmetries often seen in jets at the former wavelengths
\citep{hi94}
are mainly due to the material in the parental cloud, where the protostars 
are embedded, while observations in the radio would show symmetric jets.
In reality, this is not always the case, for instance, 
for the (high-mass) young stellar object Cepheus A HW2, 
\citet{cu06} see an asymmetry in the radio lobes and explain it by means of 
proper motions of the exciting source and
by temporal variations in the velocity of the ejections.
As we mentioned in Sect.~\ref{prop-mot}, VLA~1 does not 
show significant absolute proper motions, hence the scenario in which the 
driving source has proper motions can be ruled out. New and more sensitive 
VLA observations will be necessary to also measure proper motions of the
ejecta. Another possibility for the
ejection is due to episodic mass loss rate increments 
\citep[e.g.,][]{pe10}.
Other examples of asymmetry in the radio were presented by \citet{ro12a,ro12b} for 
the (low-mass) stars DG Tau and DG Tau B, where similar one-sided knot ejections were 
noticeable in the 3.6~cm maps.

There are two models for describing the launching region of jets: the 
disk-winds \citep{ko00} when they originate from the accretion disk and
the X-winds \citep{sh00} when they originate in the disk-star interface. 
Although a lot of progress has been done
in star-disk simulations \citep[see, e.g.,][]{pu07}, questions of whether these
winds co-exist or one 
is at work while the other is not still remain unanswered. Possible
causes for jet asymmetries may be traced back to asymmetries in the disk, e.g.,
a warped disk. Obviously, further theoretical work on the mechanisms at
work in episodic, one-sided jets is urgently needed.

Observations of VLA~1 in HH~111 together with those of DG Tau, DG Tau B, and 
Cepheus A HW2 suggest one-sided knot ejection events do occur in young stellar
objects.


\section{Summary}\label{hh111:summary}
We have performed 
      3.6~cm continuum observations with the VLA together with the analysis of additional
archival data toward the core of the HH~111 outflow. Our main results can be 
summarized as follows:
 \begin{itemize}

\item The source VLA~1 has undergone a dramatic morphological change, showing a one-sided
knot ejection at epoch 2007.

\item We have confirmed that the base of the infrared jet HH 121 is detectable at
centimeter wavelengths,
although it is weaker than the emission from the base of the optical HH 111
jet.

\item We have reported on the detection of the radio source VLA~3
located  to the north of VLA~1. However, its Galactic or extragalactic nature remains unclear.

      \item We have found that VLA~1 and VLA~3 do not show significant absolute  proper motions 
and the upper limits determined by us are consistent with the proper motions
found for (embedded) radio sources in the Orion Nebula.

   \end{itemize}

\begin{acknowledgements}
L.~G. was supported for this research through a stipend from the
International Max Planck
          Research School (IMPRS) for Astronomy and Astrophysics at
	  the Universities of Bonn and Cologne.
L.~L. and L.~F.~R. 
acknowledge the financial support of DGAPA, UNAM and CONACyT, M\'exico.
L.L. is indebted to the Alexander von Humboldt Stiftung and the Guggenheim 
Memorial Foundation for financial support. We thank the referee for providing
comments that helped to improve this paper.
\end{acknowledgements}

\bibliographystyle{rmaa} 
\bibliography{hh111b,otherb}

\end{document}